         \magnification=\magstep1

         \font\eightrm=cmr8
         \font\eightbf=cmbx8
         \font\eti=cmti8
         \hsize 12.67cm
         \def\ref{\par\smallskip\hangindent=.6cm\hangafter=1}
         \parindent .5cm
\centerline
{\bf NUCLEAR $\beta$- AND $\gamma$- COLLECTIVE BANDS IN THE
SU$_q$(2)}
\centerline {\bf ROTATOR MODEL}
\vskip .3in
\centerline{N. MINKOV, S. B. DRENSKA, P. P. RAYCHEV, R. P. ROUSSEV}
\centerline{\eti Institute for Nuclear Research and Nuclear Energy}
\centerline{\eti 72 Tzarigradsko Chaussee Blvd., 1784 Sofia, Bulgaria}
\vskip .1in
\centerline{D. BONATSOS}
\centerline{\eti European Centre for Theoretical Studies in Nuclear Physics
and Related Areas (ECT$^*$)}
\centerline{\eti Strada delle Tabarelle 286, I-38050 Villazzano (Trento),
Italy}
\vskip .3in
         \centerline{ABSTRACT}
         \vskip .1in
{\parindent 1.1cm \narrower {\parindent 0pt {\eightrm
The SU$_q$(2) rotator model is used for describing the $\beta_1$- and
$\gamma_1$-bands of even-even rare earth and actinide collective nuclei.  Good
results are obtained in nuclei with valence pair number $N>10$.  It is shown
that in the excited bands the violation of the exact SU(2) symmetry is
generally stronger than in the ground state bands, indicating the presence of
a nonadiabatic perturbation caused by the excited vibrational degrees of
freedom.  The physical content of the parameter $q$ is
discussed.  Predictions of the SU$_q$(2) model for B(E2) intraband transitions
in excited bands are presented and the need for specific experimental data is
pointed out.}}\smallskip}
\bigskip\bigskip

{\parindent 0pt {\bf 1. Introduction}}
\vskip .1in
The quantum algebra SU$_q$(2)$^{1,2}$ is a nonlinear generalization
(having the structure of a Hopf algebra$^3$) of the corresponding Lie algebra
SU(2) to which it reduces when the deformation parameter $q$ is
set equal to one. It has been found that the SU$_q$(2) algebra can be used
for describing the deviations of rotational spectra of collective
nuclei$^{4-7}$
and diatomic molecules$^{8-10}$ from the rigid rotator symmetry of
SU(2), the deformation parameter $q$ being related$^5$ to the softness
parameter of the Variable Moment of Inertia (VMI) model$^{11}$. Furthermore the
implications of the SU$_q$(2) symmetry on B(E2) transition probabilities
within the ground state bands (gsb) of deformed nuclei have been
considered$^{12}$, indicating that the B(E2) values do not saturate but
continue to increase with increasing angular momentum $I$, a result also
obtained in the framework of other models$^{13,14}$.

So far the SU$_q$(2) symmetry has been tested only in relation to levels and
B(E2) transition probabilities of the ground state band of deformed nuclei.
It is the purpose of the present work to examine the applicability of the
SU$_q$(2) symmetry to excited collective bands, the $\beta_1$- and
the $\gamma_1$-band in particular. Such an investigation is naturally
motivated by the question: "It is already known that the quantum algebra
SU$_q$(2) is appropriate to characterize nuclear rotations built on the
ground state, but what is the $q$-rotator in the case of a given excited band
where besides the rotational motion there is a presence of other collective
(vibrational) degrees of freedom?" In this respect it is interesting to study
whether the $q$-deformation "detects" the presence of the additional
nonrotational degrees of freedom. As will be seen below, the study of
the energy levels of the excited bands illuminates the above question and
leads to interesting conclusions about the physical content of the deformation
parameter $q$,
while the study of the intraband B(E2) transition probabilities emphasizes the
need for specific experimental data for testing the deviations from the pure
SU(2) behavior appearing there.
\vskip .1in
{\parindent 0pt {\bf 2. $q$-rotator definition}}
\vskip .1in
The Hamiltonian of the $q$-rotator model is proportional to the second order
Casimir operator $C_2[{\rm SU}_q(2)]$ of the quantum algebra SU$_q$(2)$^4$:
$$H={1\over {2\theta}}C_2[SU_q(2)]+E_0,  \eqno(1)$$
where $\theta$ is the moment of inertia parameter and $E_0$ is the bandhead
energy (for the gsb $E_0=0$). The corresponding energy eigenvalues are:
$$E_I={1\over 2\theta}[I][I+1]+E_0,  \eqno(2)$$
where $I$ is the angular momentum and the square brackets indicate $q$-numbers,
according to the following definition:
$$[x]={ {q^x-q^{-x}} \over {q-q^{-1}}}. \eqno(3)$$
In the case of $q$ being a phase ($q=e^{i\tau}$ with $\tau$ a real parameter),
eq. (2) gives:
$$E_I={1\over 2\theta}{\sin(\tau I)\sin\left(\tau
         (I+1)\right)\over \sin^2(\tau)}+E_0\ .\eqno(4)$$
In the limit $\tau\to 0$, the first term in eq. (4) gives the spectrum of the
usual SU(2) rigid rotator$^{15}$. It has been proved$^5$ that the deformation
parameter $\tau$ is connected to the softness parameter of the VMI model,
thus indicating that $q$-deformation is an alternative way of taking into
account nuclear stretching.
\vskip .1in
{\parindent 0pt {\bf 3. SU$_q$(2) symmetry in exited collective bands}}
\vskip .1in
In the case of excited bands one needs an appropriately formulated
$q$-rotator definition which should take into account the circumstance that
the rotational energy levels are built on a given excited vibrational
state$^{16}$.
For this purpose it is convenient to use eq. (4) in the form:
$$\overline{E_I}=E_I-E_0(I_{bh})={1\over 2\theta}{\sin(\tau
 I)\sin\left(\tau (I+1)\right)\over \sin^2(\tau)} \eqno(5)$$
 with $I>I_{bh}$, where it is supposed that the energy scale of
collective rotations has its origin in the bandhead energy $E_0(I_{bh})$, and
$I_{bh}$ is the bandhead angular momentum, which is 0 for $\beta$-bands and 2
for $\gamma$-bands. Thus, after subtracting the bandhead energy we determine
the rotational parts of the bandlevels. However, it is important to remark
that the so obtained energies are still perturbed by the vibrational motion as
far as even in the well deformed nuclei the collective rotations are not
separated completely from the vibrational degrees of freedom$^{16}$.
Taking into account this nonadiabatic perturbation we suppose that in the
excited bands the $q$-deformations of the SU(2) symmetry should be generally
larger in magnitude than the corresponding ones obtained in the gsb's. Hence
one could expect that the quantum algebraic parameter $\tau$ will be able to
indicate the presence of excited vibrational modes. Below it will be seen
that the calculations essentially support this supposition.

So, the theoretical predictions (eq. 5) are compared to the
experimental quantities $\overline{E_I^{exp}} = E_I^{exp}-E_0(I_{bh}).$ For
obtaining the fits an autoregularized iterational method of the Gauss-Newton
type$^{17}$ has been used, the quality of the fits being measured by
$$\sigma =\sqrt{ {1\over n} \sum^{I_{max}}_{I=I_{min}} (\overline{E_I^{exp}}
-\overline{E_I^{th}})^2 } , \eqno(6)$$
where $n$ is the number of levels used in the fit and $I_{min}=2$ for
$\beta$-bands, while $I_{min}=3$ for $\gamma$-bands.
We have included in the fitting procedure rare earth and actinide nuclei in
the rotational region (with  $3 \leq R_4^g = E_4/E_2 \leq 10/3$) for which
at least 5 levels of the $\beta_1$- or $\gamma_1$-band are known$^{18,19}$.
The results for the $\beta_1$- and/or $\gamma_1$-bands
of 28 rare earths and 3 actinides, along with the results
for the corresponding gsb are shown in Table 1.

The following comments can now be made:

i) The parameters $\tau_{\beta}$ and $\tau_{\gamma}$ generally obtain values
in the region 0.03--0.07, close to the typical $\tau_g$ values of 0.03--0.06
(see also $^{4-7}$). Nevertheless it is clearly seen that for almost all
considered nuclei, the $\tau$ values obtained in the excited bands lie above
the corresponding gsb values (see also Fig. 1). It turns out
that in the excited bands the quantum algebraic parameter $\tau$, which
characterizes the deviation of the spectrum from the pure SU(2)
symmetry of the rigid rotator$^{20}$, indicates the presence of
additional nonrotational degrees of freedom.  Moreover, some
$\tau_{\gamma} \geq 0.1$ values occur for nuclei with valence pair
number $N$ relatively small (10--13), indicating that the rotational
character of the $\gamma_1$-band is not yet well developed in this $N$
region.  So, the sensitivity of the SU$_q$(2) rotator description to
the structure of the different types of bands is obvious.

\par\vfill\eject

{\parindent 0pt{\eightbf Table 1.} {\eightrm Parameters of the fits of
${\beta}_1$- and ${\gamma}_1$-bands in the rare earth and actinide regions
using eq.  (5).  The deformation parameters ${\tau}_{\beta}$ and
${\tau}_{\gamma}$, the quality factors ${\sigma}_{\beta}$ and
${\sigma}_{\gamma}$ (in keV) (eq. 6) accompanied by the numbers $n_{\beta}$
and $n_{\gamma}$ of the experimental levels used in the fit, and the inertial
parameters $1/(2{\theta}_{\beta})$ and $1/(2{\theta}_{\gamma})$ (in
keV$^{-1}$) for the ${\beta}_1$- and ${\gamma}_1$-bands respectively are
shown.  The corresponding deformation parameters ${\tau}_g$ of the ground
state band and the valence pair numbers $N$ are also given.  The experimental
data are taken from $^{18,19}$.}}

         $$\vbox{\tabskip=0pt \offinterlineskip
         \halign to\hsize{\strut#\tabskip=1em plus2em
         &\hfil#&\hfil#&\hfil#&\hfil#&\hfil#&\hfil#&\hfil#&\hfil#
         \tabskip=0pt\cr
         \noalign{\hrule}
         \noalign{\vskip 2pt}
         \noalign{\hrule}
         \noalign{\vskip 2pt}
         \omit\hidewidth Nucleus\hidewidth
         &\omit\hidewidth $N$ \hidewidth
         &\omit\hidewidth ${\tau}_g$\hidewidth
         &\omit\hidewidth ${\tau}_{\beta}$\hidewidth
         &\omit\hidewidth ${\tau}_{\gamma}$ \hidewidth
         &\omit\hidewidth ${\sigma}_{\beta}[n_{\beta}]$\hidewidth
         &\omit\hidewidth $1\over 2{\theta}_{\beta}$\hidewidth
         &\omit\hidewidth ${\sigma}_{\gamma}[n_{\gamma}]$\hidewidth
         &\omit\hidewidth $1\over 2{\theta}_{\gamma}$\hidewidth \cr
         \noalign{\hrule}
$^{152}{\rm Sm}$&10&0.0622&0.0695&0.1030&22.57[8]&15.39&15.13[8] &21.63\cr
$^{154}{\rm Sm}$&11&0.0500&      &0.1306&        &     &1.14[5]  &18.67\cr
$^{156}{\rm Gd}$&12&0.0521&0.0641&0.0668&5.20[6] &12.36&11.29[10]&14.92\cr
$^{158}{\rm Gd}$&13&0.0419&      &0.1345&        &     &4.96[5]  &14.87\cr
$^{160}{\rm Gd}$&14&0.0392&      &0.0507&        &     &0.31[5]  &11.68\cr
$^{156}{\rm Dy}$&12&0.0733&      &0.0727&        &     &17.72[11]&18.66\cr
$^{160}{\rm Dy}$&14&0.0489&      &0.0715&        &     &0.71[5]  &14.24\cr
$^{162}{\rm Dy}$&15&0.0368&0.0456&0.0339&7.76[8] &8.56 &19.01[13]&12.07\cr
$^{164}{\rm Dy}$&16&0.0391&      &0.0672&        &     &3.40[5]  &11.70\cr
$^{160}{\rm Er}$&12&0.0839&      &0.1158&        &     &4.76[5]  &22.53\cr
$^{162}{\rm Er}$&13&0.0538&      &0.0605&        &     &12.19[11]&16.39\cr
$^{164}{\rm Er}$&14&0.0463&      &0.0531&        &     &19.94[13]&14.49\cr
$^{166}{\rm Er}$&15&0.0461&0.0932&0.0520&18.10[7]&12.47&5.97[13] &12.59\cr
$^{168}{\rm Er}$&16&0.0353&0.0400&0.0321&0.50[5] &9.79 &0.07[7]  &12.50\cr
$^{170}{\rm Er}$&17&0.0348&      &0.0438&        &     &10.11[6] &13.15\cr
$^{166}{\rm Yb}$&13&0.0610&      &0.0743&        &     &8.84[6]  &17.18\cr
$^{168}{\rm Yb}$&14&0.0499&      &0.0674&        &     &2.20[6]  &13.91\cr
$^{170}{\rm Yb}$&15&0.0428&0.0577&0.0342&6.49[7] &11.08&11.57[8] &13.13\cr
$^{172}{\rm Yb}$&16&0.0327&0.0584&      &7.75[8] &12.12&         &     \cr
$^{172}{\rm Hf}$&14&0.0503&      &0.0687&        &     &9.94[10] &17.55\cr
$^{174}{\rm Hf}$&15&0.0496&0.0439&      &5.07[5] &11.76&         &     \cr
$^{176}{\rm Hf}$&16&0.0449&0.0632&0.0673&3.79[6] &12.05&17.17[7] &16.26\cr
$^{178}{\rm Hf}$&15&0.0470&      &0.0867&        &     &1.90[5]  &16.22\cr
$^{180}{\rm Hf}$&14&0.0357&      &0.0434&        &     &2.24[6]  &15.15\cr
$^{178}{\rm W }$&15&0.0537&0.0545&      &6.23[8] &13.61&         &     \cr
$^{180}{\rm W }$&14&0.0591&      &0.0883&        &     &8.20[7]  &18.89\cr
$^{182}{\rm W }$&13&0.0607&      &0.1140&        &     &9.70[5]  &18.93\cr
$^{184}{\rm W }$&12&0.0476&      &0.0681&        &     &1.10[5]  &17.22\cr
$^{232}{\rm Th}$&12&0.0314&0.0378&0.0424&1.50[8] &7.07 &4.80[13] &7.44 \cr
$^{232}{\rm U }$&12&0.0364&0.0393&      &0.34[6] &7.15 &         &     \cr
$^{234}{\rm U }$&13&0.0295&0.0363&0.0514&0.45[5] &6.92 &0.29[6]  &7.12 \cr
         \noalign{\vskip 2pt}
         \noalign{\hrule}
         \noalign{\vskip 2pt}
         \noalign{\hrule}}}$$

\par\vfill\eject

ii) It is known$^{20}$ that for the ground state bands of the rare earths and
the actinides the parameter $\tau_g$ decreases with  increasing valence pair
number $N$ (or, equivalently, with increasing neutron valence pair
number $N_{\nu}$ in a given group of isotopes) approximately as
$$\tau=\sqrt{3}(8N^2+22N-15)^{-{1\over 2}}\ , \eqno(7)$$
indicating that $\tau_g$, as a measure of deviation from the rigid
rotator symmetry, indirectly reflects the nuclear shell structure.
We remark that the same trend is seen for
the $\tau_{\gamma}$ values, especially in the case of the Er isotopes
(shown in Fig. 1) and the Yb isotopes. In the excited bands it is difficult to
derive an analytical relation between $\tau$ and $N$, but Fig. 1 clearly shows
that such a correlation actually exists.  We thus conclude that in the
$\gamma$-bands the SU$_q$(2) symmetry quite well characterizes the
deterioration of the nuclear rotational properties away from the midshells.
\vskip 3.3in
{\parindent 0pt{\eightbf Fig. 1.} {\eightrm Deformation parameters ${\tau}_g$
         (circles, connected by solid lines) and ${\tau}_{\gamma}$
         (triangles, connected by dashed lines) for ground state bands and
         ${\gamma}_1$-bands respectively of Er isotopes  (taken from Table 1)
         are plotted versus the valence pair number $N$.}}
         \vskip .15in

iii) we remark that the above behavior of the parameter $\tau_{\gamma}$ allows
one to make some additional conclusions.  It has been shown$^{21}$ that in the
gsb's the correlation between $\tau$ and $N$ given approximately by eq. (7)
allows one to connect $\tau$ with the axial deformation parameter $\beta$:
$$\beta\sim{ \left ( B/[3{(2B+60.25)}^{1/2}-22.5]\right ) }^{1/2}\ ,\eqno(8)$$
where $B=1/(1-{\tau}\cot {\tau})$. Thus it has been obtained that $\beta$
decreases with the increase of $\tau$ and that $\tau$ could be considered as a
relevant measure of decrease in deformation as well as in rotational
collectivity of the nuclei in a given rotational region. Though in the excited
bands we have not such analytical estimates, Fig.  1 implies that in the
$\gamma$-bands the decrease of $\tau_{\gamma}$ towards the midshells,
could be associated similarly with the corresponding increase of nuclear
deformation and rotational collectivity.  In this case the relevance of the
quantum algebraic approach is obvious.  The data on $\beta$-bands are not
enough for drawing any conclusions about the $\tau_{\beta}$ values.
\vskip .1in
{\parindent 0pt {\bf 4. B$_q$(E2) transitions in the exited bands}}
\vskip .1in
We now turn to the study of the B(E2) transition probabilities within $\beta$-
or $\gamma$-bands. In the usual case the B(E2) values are given by
$$B(E2;I_i\to I_f)={5\over 16{\pi}}{Q_0}^2
{\vert C_{K,0,K}^{I_{i},2,I_{f}}\vert}^2,  \eqno(9)$$
where $Q_0$ is the intrinsic quadrupole moment and
$C_{m_{1},m_{2},m}^{j_{1},j_{2},j}$ are the Clebsch-Gordan coefficients of
the Lie algebra SU(2). In the case of SU$_q$(2)  one
should use the $q$-generalized angular momentum theory$^{22,23,24}$,
in which the irreducible tensor operators for the
quantum algebra SU$_q$(2)$^{23}$ as well as the
$q$-generalized version of the Wigner-Eckart theorem$^{24}$
are available.  The $q$-deformed versions of the Clebsch-Gordan
coefficients needed for the $q$-generalization of eq. (9),
         $$B_{q}(E2;I_i\to I_f)={5\over 16{\pi}}{Q_0}^2
         {\vert _{q}C_{K,0,K}^{I_{i},2,I_{f}}\vert}^2, \eqno(10)$$
are also known$^{22,23,24}$. In the case of intraband transitions with
         $\Delta I=I_i-I_f=2$ one needs
         $$_{q}C_{K,0,K}^{I+2,2,I}=
         q^{-2K}\left({[3][4][I+K+1][I+K+2][I-K+2][I-K+1]\over
         [2][2I+2][2I+3][2I+4][2I+5]}\right)^{1\over 2},  \eqno(11)$$
while in cases with $\Delta I=1$
         $${}_{q}C_{K,0,K}^{I+1,2,I} =
         -q^{I-2K+2}([I+K]-q^{2I}[I-K])  \left({[2][3][I+K+1][I-K+1]\over
         [2I][2I+2][2I+3][2I+4]}\right)^{1\over 2}  \eqno(12)$$
is needed, where the square brackets again indicate $q$-numbers
 as defined in eq. (3) with $q=e^{i\tau}$ .

Therefore in  the case of $\beta$-bands ($K=0$) one finds
$$B_{q}(E2;I+2\to I)={5\over 16{\pi}}{Q_0}^2 {[3][4][I+1]^{2}[I+2]^{2}\over
[2][2I+2][2I+3][2I+4][2I+5]}\ .\eqno(13)$$
In the case of $\gamma$-bands ($K=2$) for $\Delta I=2$ transitions  one has
$$B_{q}(E2;I+2\to I)={5\over 16{\pi}}{Q_0}^2 {[3][4][I-1][I][I+3][I+4]\over
[2][2I+2][2I+3][2I+4][2I+5]}\ ,\eqno(14)$$
while for $\Delta I=1$ transitions one finds
$$B_{q}(E2;I+1\to I)={5\over 16{\pi}}{Q_0}^2
\left([I+2]^2+[I-2]^2 -2\cos (2\tau I)[I-2][I+2]\right)$$
$$ {[2][3][I+3][I-1]\over [2I][2I+2][2I+3][2I+4]} . \eqno(15)$$
\vskip 3.3in
{\parindent 0pt {\eightbf Fig.  2.} {\eightrm $B_{q}(E2;I+2\to I)$
transition probabilities are plotted as a function of angular momentum $I$
in the cases of $\beta$-bands (eq. 13, solid lines) and $\gamma$-bands
(eq. 14, dashed lines) for some typical values of the deformation
parameter $\tau$.  The numerical values of $B_{q}(E2)$ correspond to
${5\over 16\pi}Q_0^2=1$.  The limiting case $\tau=0$ gives the usual rigid
rotator predictions.}}
\vskip .15in
On these results the following comments apply:

i) eq. (13), concerning the $\beta$-bands,
 is exactly the same as the one obtained in the case of gsb$^{4,12}$.
 It has been shown that this equation gives B(E2) values increasing
with increasing $I$, while the corresponding usual SU(2) expression
(obtained here for $\tau \to 0$) exhibits saturation with increasing $I$.
This is illustrated in Fig. 2. In the case of gsb's some experimental examples
supporting this prediction have been given in $^{12}$. Similar predictions
also occur in the framework of other models$^{13,14}$.
The existing data for
$\beta_1$-bands do not suffice for testing this prediction.

ii) eq. (14), concerning $\Delta I=2$ transitions in $\gamma$-bands, gives
almost the same behavior as eq. (13), as seen in Fig. 2. It follows that
for $\Delta I=2$ transitions the introduction of $q$-generalized
Clebsch-Gordan coefficients leads to a typical modification of the reduced
transition probabilities in all considered bands.
\vskip 3.3in

{\parindent 0pt {\eightbf Fig. 3.} {\eightrm Same as Fig. 2 but for the
case of $B_{q}(E2;I+1\to I)$ transition  probabilities in $\gamma$-bands.}}
\vskip .15in
iii) eq. (15), concerning $\Delta I=1$ transitions in $\gamma$-bands,
illustrated in Fig. 3, gives an interesting prediction.
For typical $\tau$-values (0.03--0.07) one initially observes a decrease
of $B_{q}(E2;I+1\to I)$ with increasing $I$, but further, after reaching some
minimum (for example at $I=5$ when $\tau =0.05$), a significant
increase of  $B_{q}(E2)$ is observed, while in the rigid rotator limit
($\tau\to 0$) a continuous decrease down to zero at sufficiently large
$I>12$ is predicted. The available data for
$E2$ intraband transitions in the excited bands do not suffice for
 detailed tests of these predictions,  due to the short life times
and strong $M1$ mixing observed in these transitions. The need for further
experimental data is clear. In particular  the observation of any $E2$
transitions with $\Delta I=1$ at $I>10-12$ in the
$\gamma$-bands will be useful in testing the predictions of eq. (15).

We now remark that the present investigation outlines the principal limits of
the SU$_q$(2)-symmetry approach to the nuclear rotational spectra. It should
be emphasized that in the framework of the quantum algebra SU$_q$(2) as well
as in the case of the standard Lie algebra SU(2)$^{25}$, one is able to
provide a consistent description of the physical characteristics of only one
given rotational band. This is clearly indicated by the distinctions in the
magnitudes of the $q$-deformation parameter obtained for the different types
of bands (see Table 1). It follows that one should understand the
SU$_q$(2)-rotator as a one-band model based on the particular intrinsic
state or vibrational mode. Hence the unified description of the different
rotational bands including the calculation of the interband transition
probabilities is beyond the limits of the quantum algebra SU$_q$(2). Such an
extension could be referred to a model based on the $q$-deformed algebra
SU$_q$(3) in which the introduction of a bandmixing interaction would be
possible. However the realization of such a model is still complicated due to
some difficulties in the obtaining of the reduction SU$_q$(3)$\supset$SO$_q$(3)
(for example see $^{26,27}$). In this respect the use of the
simplest quantum algebra SU$_q$(2) could be considered as a first
approximation in the construction of a more complicated quantum algebraic
theory of nuclear collective motion.
\vskip .1in
{\parindent 0pt {\bf 5. Conclusion}}
\vskip .1in
In conclusion, we have demonstrated the relevance of the SU$_q$(2) approach
beyond the ground state bands, namely in the excited bands of even-even rare
earth and actinide nuclei.  Good results have been obtained for $\beta_1$ and
$\gamma_1$ bands in nuclei with valence pair number $N>10$.  The quantum
algebraic parameter $\tau$ fitted in these bands obtains values generally
shifted above the corresponding ones in the gsb's.  In such a way the
$q$-deformation specifically indicates the presence of a nonadiabatic
perturbation caused by the excited vibrational degrees of freedom.
The decrease of $\tau_{\gamma}$ and $\tau_g$ with increasing
$N$ is in accordance with the interpretation of $\tau$ as a measure of
deviation from the rigid rotator limit equivalent to the nuclear
softness$^{5,20}$.
In addition, these correlations (illustrated in Fig.  1) allow one to
extend the SU$_q$(2) symmetry to a wider range of nuclear rotational
properties$^{21}$.  The predictions of the SU$_q$(2) rotator model for the
$B(E2)$ intraband transition probabilities in the excited bands show
modifications in comparison to the SU(2) rigid rotator limit, the experimental
data needed for testing these predictions having been identified.  It is
pointed out that SU$_q$(2) is a simple one-band model, but it can be
considered as the first step in the development of more complicated models
based on the $q$-deformed algebras.
\vskip .1in
\parindent=0pt
{\bf 6. References}
\vskip .1in
\ref {1.}\ L. C. Biedenharn, {\it J. Phys.} {\bf A 22} (1989) L873.

\ref {2.}\ A. J. Macfarlane, {\it J. Phys.} {\bf A 22} (1989) 4581.

\ref {3.}\ V. G. Drinfeld,  in {\it Proceedings of the  International
         Congress of Mathematicians}, edited by  A. M. Gleason
         (American Mathematical Society, Providence, RI, 1987) p. 798.

\ref {4.}\ P. P. Raychev, R. P. Roussev and Yu. F. Smirnov, {\it J.
         Phys.} {\bf G 16} (1990) L137.

\ref {5.}\ D. Bonatsos, E. N. Argyres, S. B. Drenska, P. P.
         Raychev, R. P. Roussev and Yu. F. Smirnov, {\it Phys. Lett.} {\bf
         B251} (1990) 477.

\ref {6.}\ S. Iwao, {\it Prog. Theor. Phys.} {\bf 83} (1990) 363.

\ref {7.}\ E. Celeghini, R. Giachetti, E. Sorace and
         M. Tarlini, {\it Phys. Lett.} {\bf B280} (1992) 180.

\ref {8.}\ D. Bonatsos, P. P. Raychev, R. P. Roussev and Yu. F. Smirnov,
{\it Chem. Phys. Lett.} {\bf 175} (1990) 300.

\ref {9.}\ Z. Chang and H. Yan, {\it Phys. Lett.} {\bf A154} (1991) 254.

\ref {10.}\ J. G. Esteve, C. Tejel and B. E. Villaroya, {\it J. Chem. Phys.}
{\bf 96} (1992) 5614.

\ref {11.}\ M. A. J. Mariscotti, G. Scharff-Goldhaber and B. Buck, {\it Phys.
Rev.} {\bf 178} (1969) 1864.

\ref {12.}\ D. Bonatsos, A. Faessler, P. P. Raychev, R. P. Roussev and Yu. F.
Smirnov, {\it J. Phys.} {\bf A 25} (1992) 3275.

\ref {13.}\ M. Mukerjee, {\it Phys. Lett.} {\bf B251} (1990) 229.

\ref {14}.\ J. L. Pin, J. Q. Chen, C. L. Wu and D. H. Feng, {\it Phys. Rev.}
{\bf C 43} (1991) 2224.

\ref {15.}\ A. Bohr and B. R. Mottelson, {\it Nuclear Structure} Vol. II
(Benjamin, New York, 1975).

\ref {16.}\ J. Eisenberg and W. Greiner, {\it Nuclear Theory} Vol. I
(North-Holland, Amsterdam, 1970).

\ref {17.}\ L. Aleksandrov, {\it Math. Phys. Comput. Math.} {\bf 11} (1971) 36.

\ref {18.}\ M. Sakai, {\it At. Data Nucl. Data Tables} {\bf 31} (1984) 399.

\ref {19.}\ V. M. Belenki and E. P. Grigoriev, {\it Structure of
         Even Nuclei} (Energoatomizdat, Moscow, 1987).

\ref {20.}\ N. Minkov, R. P. Roussev and P. P. Raychev,
         {\it J. Phys.} {\bf G 20} (1994) L67.

\ref {21.}\ N. Minkov, P. P. Raychev and R. P. Roussev,
         {\it J. Phys.} {\bf G 21} (1995) 557.

\ref {22.}\ Yu. F. Smirnov, V. N. Tolstoy and Yu. I. Kharitonov, {\it Yad. Fiz.}
{\bf 53} (1991) 959 [{\it Sov. J. Nucl. Phys.} {\bf 53} (1991) 593];
{\it Yad. Fiz.} {\bf 53} (1991) 1746 [{\it Sov. J. Nucl. Phys.} {\bf 53}
     (1991) 1068].

\ref {23.}\ F. Pan, {\it J. Phys.} {\bf A 24} (1991) L803.

\ref {24.}\ A. U. Klimyk, {\it J. Phys.} {\bf A 25} (1992) 2919.

\ref {25.}\ E. Chac\'on, M. Moshinsky and V. Vanagas,
     {\it J. Math. Phys.} {\bf 22} (1981) 605.

\ref {26.}\ C. Quesne, {\it Phys. Lett.} {\bf B304} (1993) 81.

\ref {27.}\ J. Van der Jeugt, {\it J. Math. Phys.} {\bf 34} (1993) 1799.

         \vfill\eject\bye